\documentclass[prl,showpacs,amsmath,amssymb,preprint,endfloats]{revtex4}

\usepackage{graphicx}
\usepackage{dcolumn}
\usepackage{bm}

\begin{document}

\preprint{Draft}

\title{XMCD characterization of rare-earth dopants in Ni$_{81}$Fe$_{19}$(50nm): \\microscopic basis of engineered damping }

\author{W. E. Bailey\email{web54@columbia.edu}, H. Song, and L. Cheng}
\affiliation{Materials Science Program, Department of Applied
Physics, 200 S.W. Mudd Bldg, Columbia University, New York, NY
10027}

\date{\today}

\begin{abstract}

We present direct evidence for the contribution of local orbital
moments to the damping of magnetization precession in magnetic
thin films.  Using x-ray magnetic circular dichroism (XMCD)
characterization of rare-earth (RE) M$_{4,5}$ edges in
Ni$_{81}$Fe$_{19}$ doped with $<$ 2\% Gd and Tb, we show that the
enhancement of GHz precessional relaxation is accompanied by a
significant orbital moment fraction on the RE site.  Tb
impurities, which enhance the Landau-Lifshitz(-Gilbert) LL(-G)
damping $\lambda(\alpha)$, show a spin to orbital number ratio of
1.5$\pm$0.3; Gd impurities, which have no effect on damping, show
a spin to orbital number ratio of zero within experimental error.
The results indicate that the dopant-based control of
magnetization damping in RE-doped ferromagnets is an atomistic
effect, arising from spin-lattice coupling, and thus scalable to
nanometer dimensions.

\end{abstract}

\maketitle

\section{Introduction}

Ultrafast magnetization dynamics are important for determining the
data rate ($>$ 1GHz) of magnetic information storage.\cite{koch}
Recent advances in time-domain measurement techniques have
provided clear pictures of the ensemble\cite{russek1,pimm} and
sub-micron spatial\cite{back} response of magnetization to short
magnetic field pulses, governed by the Landau-Lifshitz (LL) or
-Gilbert (LLG) equations. The LL relaxation rate $\lambda$ or
Gilbert damping $\alpha$ determines the characteristic time for
the magnetization to relax into new equilibria.  Characteristic
times are given by $2/\lambda$ (LL) or $2 /
\mu_{o}M_{s}\gamma\alpha$ (LLG) and are on the order of 2 ns in
Ni$_{81}$Fe$_{19}$(50nm).

Dilute concentrations of rare-earth (RE( dopant atoms can be used
to increase the damping of the precession in Ni$_{81}$Fe$_{19}$,
speeding the return of magnetization to
equilibrium.\cite{bailey-Tb,ingvarsson} Terbium (Tb, 0-10\%) has
been shown to provide roughly two orders of magnitude of
enhancement in $\alpha$\cite{bailey-Tb}, from
$\textrm{0.006}\leq\alpha\leq\textrm{0.7}$. The contributed
damping has been found to be general for lanthanide dopants Sm-Ho,
with effectiveness scaling roughly with the nominal orbital moment
$<\hat{L_{RE}}>$ of the dopant $4f$ shell.\cite{reidy-apl}.
Consistent with this idea, and with early measurements of FMR
linewidth in RE-substituted YIG\cite{kittel-fast,seiden}, Gd
dopants have shown no significant effect on damping.

Direct evidence has not been available previously to link RE
impurity magnetization states to precessional damping, either in
YIG or in modern thin-film magnetic systems. The rare earth
magnetization is usually approximated as that of an isolated $4f$
shell, occupied as $4f^{Z-57}$, where $Z$ is the atomic number of
the RE impurity; spin, orbital, and total moments ($S,L,J$) are
calcuated using Hund's rules.  This approach yields moments which
agree well with experimental moments except for metallic
lanthanides near Eu\cite{chikazumi}; measurement of the magnetic
character of Gd and Tb in a metallic alloy is therefore
worthwhile.

X-ray magnetic circular dichroism (XMCD) is an ideal tool to
characterize the magnetic character of RE dopants.  Orbital and
spin moments $<L>$ and $<S>$ can be measured separately on
individual atomic sites using sum
rules\cite{tholePRL1992,tholePRL1993}. High resolution XMCD data
have been measured previously in Tb single crystals at M$_{4,5}$
edges\cite{brookes-tb} and in Gd single crystals at the M$_{5}$
edge\cite{brookes-gd}.  To our knowledge, however, sum rules have
not yet been applied to the dilute RE impurities in transition
metal ferromagnets relevant for controlled damping.

We have used XMCD to measure spin to orbital moment ratios of Gd
and Tb (2\%) in Ni$_{81}$Fe$_{19}$(50nm).  We show that for these
elements, the calculated $L/S$ ratios from a $4f^{Z-57}$ shell are
verified.  A fivefold enhancement in GHz relaxation rate from Tb
dopants is accompanied by a large orbital moment fraction on the
Tb site, indicating that spin-lattice coupling is decisive in
enhancing relaxation.

\section{Experimental}

Films were deposited and magnetization dynamics were measured
using methods described in ref. \cite{reidy-apl}.
Ni$_{81}$Fe$_{19}$ (50 nm) thin films were prepared with 2\%
atomic concentrations of Gd and Tb.  Films were deposited using
ion beam deposition in a load-locked, multitarget chamber with
base pressure of 6x10$^{-8}$ Torr.  Doping concentrations were
measured as 1.7\% of Gd and 1.8\% of Tb using Rutherford
Backscattering Spectroscopy (RBS).  The top surface of the films
was protected by a 20 {\AA} Ta cap layer\footnote{This thickness
was found to be optimal for XMCD measurements, as zero cap layer
thickness produced Fe and Ni MCD spectra characeristic of oxide,
and 50 {\AA} Ta caps provided a large attenuation of the MCD
signal.}

Magnetization dynamics of the thin films were characterized using
time-domain pulsed inductive microwave magnetometry (PIMM).
Measured waveforms are proportional to $\partial\phi/\partial
t(t)$, where $\phi$ is the in-plane angle of the magnetization.  A
bias field H$_{B}$ is applied along the magnetic (induced)
easy-axis direction, orthogonal to the pulsed field; $H_{B}=$ 20
Oe unless noted otherwise.  See refs \cite{pimm,bailey-Tb} for
details. Magnetization dynamics were measured within one day of
deposition; samples were stored in a dessicator for less than a
week before XMCD measurement.

XMCD measurements were taken in total electron yield mode (TEY) at
the UV ring of the National Synchrotron Light Source (NSLS),
Brookhaven National Laboratory, Beamline U4B.\cite{idzerda-ty-tey}
XMCD was measured for fixed circular photon helicity, 75\%
polarization, with pulsed magnetization switching
($H=\pm\textrm{300 Oe}$) at the sample; photon incidence was fixed
at 45$^{\circ}$ with respect to the sample normal.  The samples
were mounted with magnetic easy axis along the applied field
direction; measurements taken at remanence and in a saturating
field were not found to differ appreciably.  XMCD measurements
with $H+(M+)$ measured first and $H-(M-)$ measured second were
averaged with measurements taken in reversed order ("duplex
mode"), to correct for any drift in the monochromator which might
lead to derivative-like artifacts in XMCD.  XMCD measurements were
divided by the factor $0.75\cos{\textrm{45}^{\circ}}$ to correct
for non-grazing incidence and incomplete circular polarization.

Sample TEY currents were normalized to TEY currents measured at a
reference grid (I$_{o}$), located ahead of the sample, to correct
for any variation in beam intensity over the measurement.  Photon
energies could be varied continuously in the experiment from 500 -
1350 eV using a grating monochromator, with a general drop in beam
intensity towards higher photon energies. High photon energies
(1130-1320 eV) were calibrated using electron yield signals from
inline Eu$_{2}$O$_{3}$ and Dy$_{2}$O$_{3}$ powder references,
taking RE edge positions for the oxides reported in ref.
\cite{kaindl-jap}. A +3kV extraction voltage was applied near the
sample surface; this was found to be important for reproducible
MCD difference spectra.

\section{Results}

Ultrafast magnetization dynamics measurements, taken by pulsed
inductive microwave magnetometery (PIMM), are shown in Figure 1.
The responses of three films are shown, for bias fields H$_{B}$=20
Oe: undoped Ni$_{81}$Fe$_{19}$, Gd-doped Ni$_{81}$Fe$_{19}$, and
Tb-doped Ni$_{81}$Fe$_{19}$, each 50 nm thick with a 2 nm Ta cap.
The fast risetime pulse is applied at $t\simeq\textrm{0.7ns}$.

It can be seen that the Tb-doped sample experiences a much larger
damping of magnetization motion than do the undoped or Gd-doped
samples. This behavior can be connected with the materials
parameter $\lambda$ through the LL equation (SI units),

\begin{equation}
{d\mathbf{M}\over dt} = - \mu_{0}\mid \gamma \mid \left(\mathbf{M}
\times \mathbf{H}\right) - {\lambda\over M_{s}^{2}}
\left(\mathbf{M} \times \mathbf{M} \times \mathbf{H} \right),
\label{LLG}
\end{equation}

where $\mathbf{M}$ is the magnetization, $\mathbf{H}$ is the
effective applied field including demagnetizing and anisotropy
components, $\gamma$ is the gyromagnetic ratio, and $\lambda$ is
the relaxation rate in $s^{-1}$.  The second term describes the
relaxation of the motion.  A time-domain solution can be written
valid for small rotation angles and single domain
behavior\cite{silva1}

\begin{equation}
\large
\phi(t)=\phi_{0}+\beta_{0}e^{-\lambda\:t/2}\sin{\left(\omega_{p}
t+\varphi\right)}
\end{equation}

where $\phi_{0}$, $\beta_{0}$, and $\varphi$ are constants.  Fits
to this equation are pictured with the PIMM data.  We find
relaxation rates $\lambda$ of 0.9 GHz for undoped
Ni$_{81}$Fe$_{19}$, 0.90 for Gd(2\%)-doped Ni$_{81}$Fe$_{19}$, and
4.2 GHz for Tb(2\%)-doped Ni$_{81}$Fe$_{19}$.

\begin{figure}[h]
\includegraphics[width=\columnwidth]{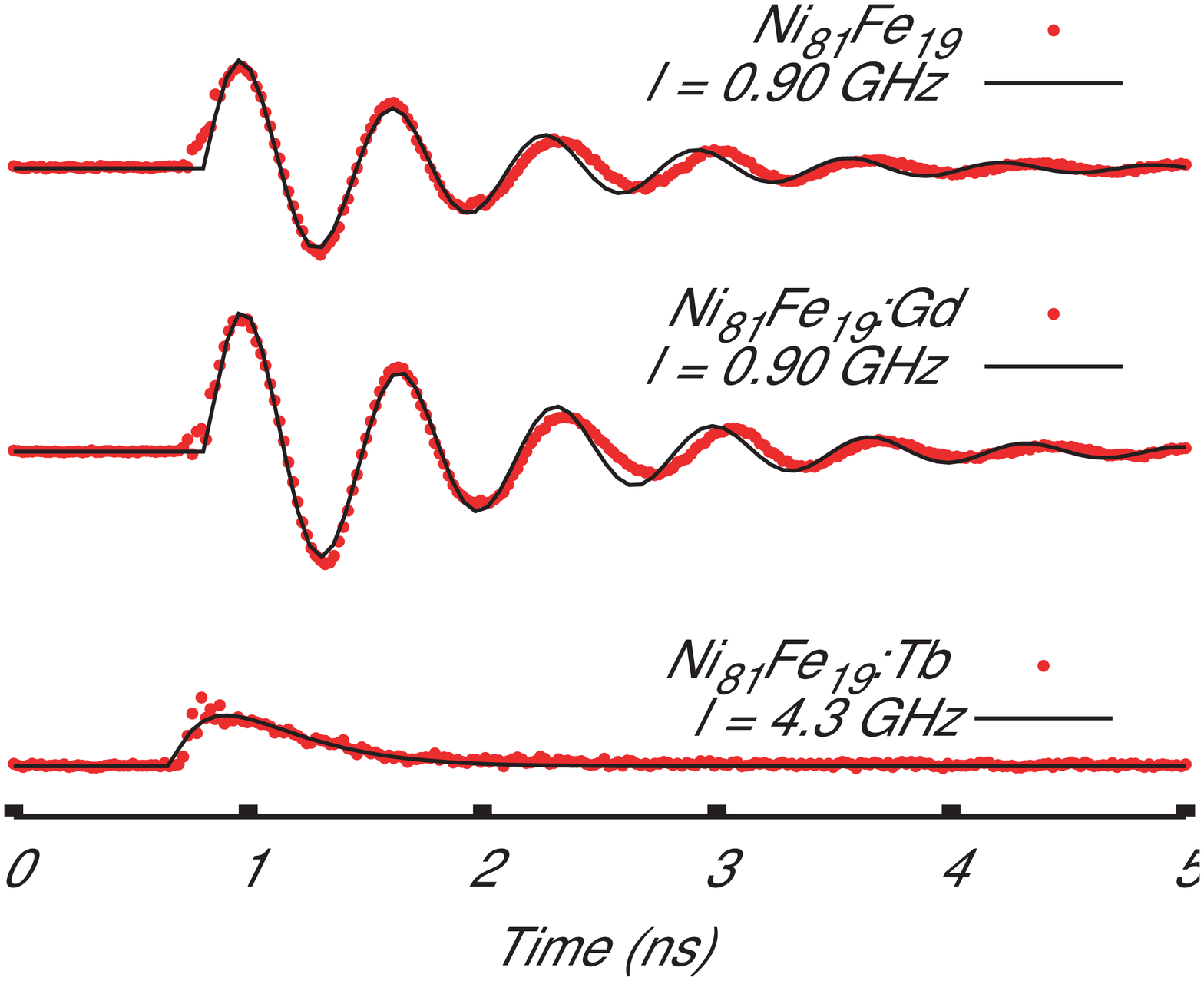}
\caption{Inductive measurement of magnetization dynamics for
undoped, Gd(2\%)-doped, and Tb(2\%)-doped Ni$_{81}$Fe$_{19}$(50nm)
thin films, with LL-model fits.}
\end{figure}

\begin{figure}[h]
\includegraphics[width=\columnwidth]{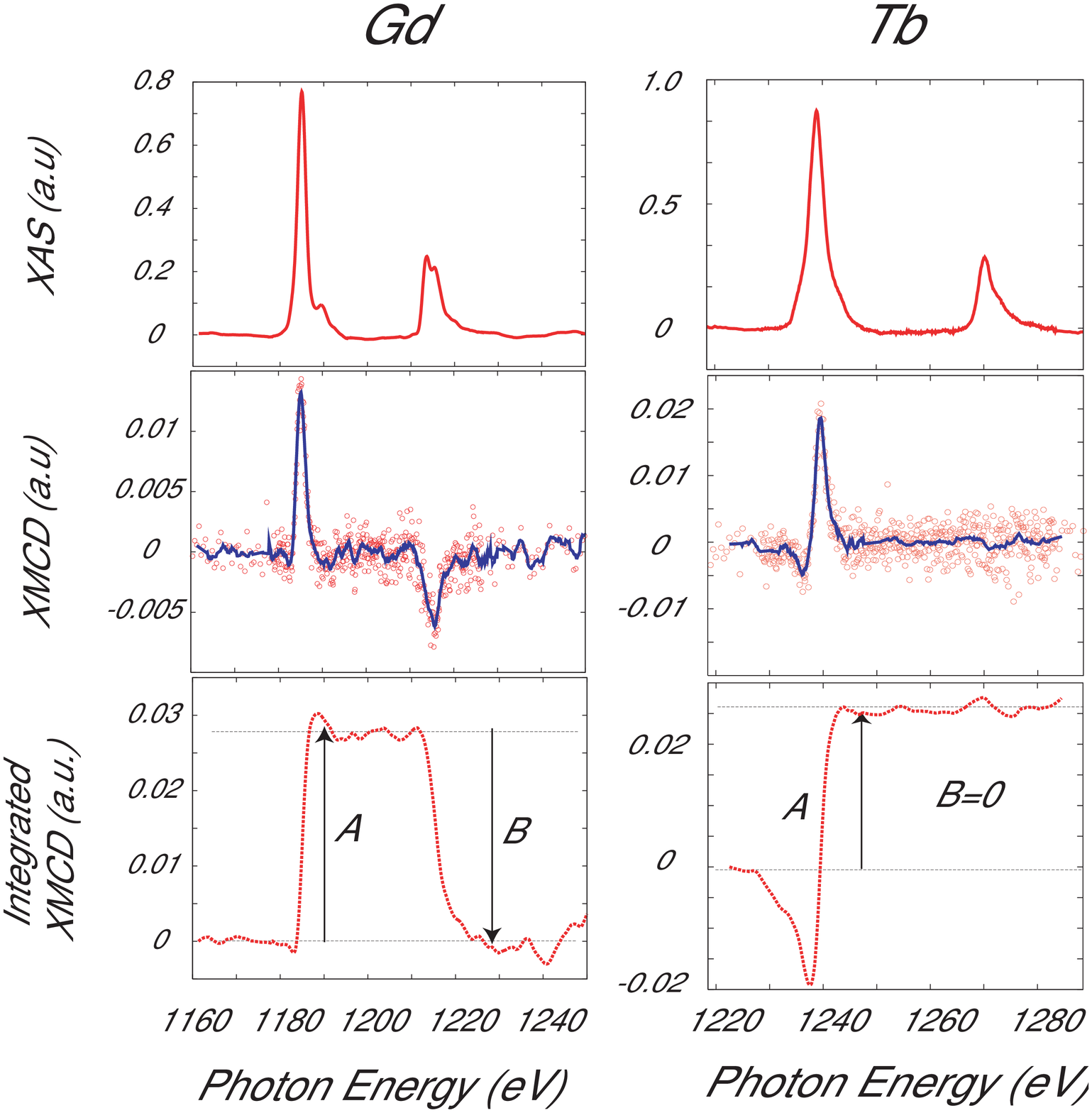}
\caption{XMCD characterization of RE dopant magnetism in
Ni$_{81}$Fe$_{19}$:RE(2\%), RE=Gd,Tb. {\it Top:} x-ray absorption
spectra (XAS), {\it middle:} XMCD difference data (circles) with
polynomial smoothing fit (lines); {\it bottom:} energy integral of
XMCD spectra.  Bottom figures are the integrals of middle figures.
The M$_{5}$ XMCD integral $A$ and M$_{4}$ XMCD integral $B$ are
indicated.} \label{xmcd}
\end{figure}

XMCD characterization of the local magnetization states on RE
dopants is shown in Figure \ref{xmcd}.  X-ray absorption spectra
for Gd and Tb are shown in the top panel; XMCD measurements are
shown in the middle panel (dots).  The XMCD data have been
smoothed using a polynomial fit with variable window position
(lines); the smoothed data, integrated numerically over energy,
are shown in the bottom panel.

The nominal XAS peak positions at M$_{5}$ and M$_{4}$ correspond
within $\pm\textrm{0.7 eV}$ to published values for elemental Tb
and Gd samples\cite{brookes-gd,brookes-tb,thole-lanthanides}. Main
peaks are seen for Gd at 1184.7 eV(M$_{5}$), 1213.8 eV(M$_{4}$)
and Tb at 1238.7 eV(M$_{5}$), 1271.7 eV(M$_{4}$). Additionally, Gd
exhibits a small shoulder on the high energy side of M$_{5}$ and a
split peak in M$_{4}$ consistent with
refs\cite{brookes-gd,thole-lanthanides}.

XMCD characteristics are quite different for Gd and Tb.  Gd shows
a positive and negative peak at M$_{5}$ and M$_{4}$ respectively.
Tb shows a net positive peak at M$_{5}$, with a small negative
peak on the low energy side, and a small disturbance at $M_{4}$ on
the threshold of experimental error.  Tb observations are
consistent with elemental Tb XMCD spectra found by van der Laan et
al\cite{brookes-tb}; here the negative peak on M$_{5}$ and small
positive and negative peaks at M$_{4}$ are roughly 15\% and 7\%
the heights of the large positive peak at M$_{5}$.

The bottom panel shows the energy integrals of the two XMCD peaks,
following the method in \cite{chen} for application of sum rules.
$A$ and $B$ are the integrals over $M_{5}$ and $M_{4}$ XMCD,
respectively.  Orbital to spin number ratios $<L>/<S>$ obtained at
M$_{4,5}$ edges can be determined from the
formula\cite{tholePRL1992,tholePRL1993}

\begin{equation}
{<\hat{L}_{z}>\over <\hat{S}_{z}>}=2{|A|-|B|\over |A|+{3\over
2}|B|}\left(1+{<\hat{T}_{z}>\over <\hat{S}_{z}>}\right)
\end{equation}

where $T_{z}$ is the magnetic dipole operator.  We extract, for
Gd, $A=B=\textrm{0.027}\pm\textrm{0.004}$, and for Tb,
$A=\textrm{0.026}\pm\textrm{0.002}$,
$B=\textrm{0.000}\pm\textrm{0.002}$.  Jo et al\cite{jo-sum-rule}
have published estimates for $<T_{z}> and <S_{z}>$ for all
lanthanides, excepting Eu, according to atomic calculations,
yielding $<T_{z}/S_{z}>_{Gd}=-\textrm{0.009}$ and
$<T_{z}/S_{z}>_{Tb}=-\textrm{0.08}$.

Estimates of orbital to spin number ratios (twice the magnetic
moment ratios) are given in Table \ref{resulttable}.  We find
$<L_{z}/S_{z}>_{Gd}=\textrm{0.0}\pm\textrm{0.1}$, and
$<L_{z}/S_{z}>_{Tb}=\textrm{1.5}\pm\textrm{0.3}$.  These values
are in qualitative agreement with Hund's rule estimates.  Gd
(Z=64, $4f^{7}$) has $S=7/2$, $L=0$ ($L/S=\textrm{0}$), and Tb
(Z=65, $4f^{8}$) has $S=3$, $L=3$ ($L/S=\textrm{1}$).

\begin{table}
  \centering
  \begin{tabular}{|c|c|c|}
    \hline
     L/S ratio& Tb & Gd \\
    \hline
    XMCD experiment & 1.5$\pm$0.3 & 0.0 $\pm$ 0.1\\
    Hund's rule, $4f^{Z-57}$ & 1 & 0 \\
    \hline
  \end{tabular}
\caption{Orbital to spin moment ratios $L/S$ on rare-earth dopants
RE=(Tb, Gd) in Ni$_{81}$Fe$_{19}$:RE(2\%)(50nm), as measured by
XMCD and predicted by Hund's rules.  The range of $L/S$ estimate
for Tb depends on the spin sum rule used; see text for
details.}\label{resulttable}
\end{table}

\section{Discussion}

XMCD measurements show a great difference between the magnetic
character of dilute Gd and Tb in Ni$_{81}$Fe$_{19}$.  Gd is found
to be pure spin type ($S$-state) and dilute Tb is found to have
roughly equal parts spin and orbital moment.   The approximation
of the isolated $4f^{Z-57}$ moment is broadly validated for these
two rare-earth dopants in Ni$_{81}$Fe$_{19}$, although the Tb
$L/S$ value is roughly 50\% higher than that found through
calculation. Alternate handling of the spin sum rule, such as the
conventional neglect of the $<T_{z}>$ term, increases the
disagreement.  Based on the conclusion of \cite{jo-sum-rule}, the
validity of the spin sum rule is not seriously in question for
these two elements, although it may not hold for the lighter
lanthanides.

The XMCD measurement verifies an important criterion for an
atomistic basis of contributed damping in doped
Ni$_{81}$Fe$_{19}$.  It has been proposed that the presence of
spin-orbit coupling is essential for the damping of uniform
precession by electronic
excitations\cite{korenman-prange,kambersky-microscopic} which can
ultimately be absorbed by a phonon and dissipated as heat.
Rare-earth elements can provide local centers for spin-orbit
coupling: the orbital moment of the RE can couple to the Fe, Ni
spin system through the RE spin moment.  A necessary criterion for
this mechanism is the presence of an orbital moment on RE sites
which enhance the damping.  We have validated its presence in Tb
and absence in Gd.

\section{Conclusion}

We have seen that XMCD characterization of Tb and Gd dopants in
Ni$_{81}$Fe$_{19}$ reveals a large orbital moment fraction on Tb
sites, accompanied by a large increase in precessional damping,
but zero orbital moment on Gd, with no effect on precessional
damping.  The results provide support for the idea that spin-orbit
coupling, through introduction of local orbital moments, is
important for controlled damping from lanthanide dopants.

\section{Acknowledgements}

We thank Dario Arena and Joe Dvorak (NSLS / U4B) for beamline
support and Sasha Bakru for RBS measurements. Research was carried
out in part at the National Synchrotron Light Source, Brookhaven
National Laboratory, which is supported by the U.S. Department of
Energy, Division of Materials Sciences and Division of Chemical
Sciences, under Contract No. DE-AC02-98CH10886.  We acknowledge
the NIST Nanomagnetodynamics Program (606NANB2D0145) for support.


\end{document}